\documentclass[preprint]{elsarticle}

\usepackage[T1]{fontenc}

\biboptions{sort&compress}

\usepackage{lineno,hyperref}
\modulolinenumbers[5]

\begin{document}

\begin{frontmatter}

\title{Size effects in stress penetration and dynamics of dislocations: Fe-Ni-Cr steel}

\author{Zbigniew Kozio{\l}}
\address{National Center for Nuclear Research, Materials Research Laboratory\\ul. Andrzeja So{\l}tana 7, 05-400 Otwock-{\'S}wierk, Poland}
\ead{zbigniew.koziol@ncbj.gov.pl}

\begin{abstract}
Movement of edge (line) dislocations in FCC steel 310S is shown to depend on the size of nanoscale structures, based on modeling withing molecular dynamics (MD). The effect is attributed to time (and size) dependencies of pressure propagation into the medium interior. The observation is crucial in interpreting any  MD studies of pressure effects since these are governed by time-dependent internal virial stresses. In particular, velocity of dislocations scales well with value of local internal shear component of virial stress $S_{xy}$ and not with external shear pressure. Dynamics of stress penetration is described well within the model of damped harmonic oscillator, where characteristic oscillation frequency depends on number of crystallographic layers in direction along the wave propagation while the speed of stress propagation is the speed of sound. The minimal stress required for dislocation movement (Peierls stress) is determined to be 0.75 GPa. Pressure and temperature effects on dislocation movement are systematically investigated.

\end{abstract}



\end{frontmatter}


\section{Introduction}

\subsection{Continuum interpretation of virial stress in molecular simulations.}

The multiscale constitutive modeling numerical simulations are playing an important role in the understanding of the fundamental mechanisms governing microstructural evolution. The problem of scale bridging between Molecular Dynamics and Continuum Mechanics analyses remains challenging, hindering the simultaneous study of physical processes at the atomic and continuum levels, especially in irradiated systems. 

In the irradiated materials, all the processes take place during and soon after the interaction of energetic incident particles with lattice atoms. These processes are experimentally unobservable because a displacement phase of the collision cascade usually lasts picoseconds. The only approach that may address this issue is the use of Molecular Dynamics simulations. 

The efforts to develop validated methodologies capable of predicting microstructural evolution and mechanical property changes are hampered by problems associated with the stress definition and its dynamics, applicable to both continuum and discrete systems. 

Stress is one of the most fundamental quantities in Continuum Mechanics, it is essential to introduce a stress definition applicable to both continuum and discrete systems. The most common one in discrete-particle systems is based on virial theorem. The virial stress consists of at least two components: a kinetic component depending on the atomic particle mass and velocity and a potential component depending on the interatomic forces and atom positions:

\begin{equation}
    S_{ij}=\frac{1}{V} \sum_{k \in V}\left( -m^k\left(v^k_iv^k_j\right)+\frac{1}{2} \sum_ {l \in V}\left(x^l_i-x^k_i\right)F^{kl}_j\right),
\label{eq:virial}\end{equation}

where $k$ and $l$ indexes atoms in the volume $V$, $m^k$ is the mass of atom $k$,$v^k_i$, $x^l_i$ is the $i$th 
component of the velocity and position of atom, while $F^{kl}_j$ is the force along direction $j$ on atom $k$ due to atom $l$.

It has been reasoned that the virial stress represents an atomistic definition of stress that is equivalent to the continuum Cauchy stress \cite{Subramaniyan}, provided spatial and temporal averages are computed properly \cite{Zimmerman}, \cite{Elder}, while the problem attracted a lot of discussion \cite{Yang}, \cite{Zhou}.

\subsection{Stress propagation into a medium.}
\label{stress-sound-propagation-into-a-medium}

There are two basic approaches towards modeling of stress effects in materials with the use of MD \cite{Simar}. Either an external pressure is applied and its effects are studied as a function of time (so called dynamical simulations) or a continuous, changing linearly in time ramp force is applied (so called quasi-static simulations). In any of these cases we deal in fact with a dynamic situation and any interpretation of computer modeling results must be based on understanding the time scales involved and spacial distribution of stress penetration. 

Longitudinal \(c_L\), and transversal \(c_T\) speed of stress (sound) propagation is given by (see e.g. \cite{Kinsler}): 
$c_L = \left(E/\rho\right)^{1/2}$, and $c_T = \left(G/\rho\right)^{1/2}$,
where \(E\) and \(G\) is Young modulus and shear modulus, respectively, and \(\rho\) is material density.

For finding elastic constants we used a package written by Aidan Thompson available in source code distribution of LAMMPS, which is based on the modeling concept of Sprik et al. \cite{Sprik}. At low temperatures for Bonny potential \cite{Bonny} elastic constants, $C_{11}$, $C_{12}$ and $C_{44}$, are, respectively, 327.06, 189.27 and 156.9 GPa. Poisson ratio obtained is 0.36 (the result closest to experimental one for all potentials studied). At T=300K density is 8.09 $g/cm^3$, which corrected for real atomic content of 310S by a factor 0.989 gives the density 7.999 $g/cm^3$ at 300K. 

With computed elastic parameters we find the values of \(c_L\) and \(c_T\) at T=0K: \(c_L\) is 6308 m/s and \(c_T\) is 4368 m/s. Both, \(c_L\) and \(c_T\) almost do not depend on temperature and are of the order of 63 {\AA}/ps and 44 {\AA}/ps,  in a reasonable agreement with known values for steel and iron \cite{Decremps}. 

We distinguish between microscopic pressure tensor components as reported by the MD modeling software LAMMPS \cite{LAMMPS}\footnote{https://www.lammps.org/} and pressure values applied at the surface,  which is related to surface average of forces. Additionally, we ought to distinguish between the pressure reported by LAMMPS at the surface, and the external pressure applied to that surface. The last one will be denoted with superscript 0. Hence, $P^0_{xy}(t)$, the applied shear pressure caused by force in X-direction on surface Y is in general not the same as the pressure exerted on the surface by sample interior, $P_{xy}(t)$, at any given moment of time. 

When pressure is applied abruptly at the surface in a form of Heaviside function, \(P^0_{xy}\) =0 for t\textless{}0 and \(P^0_{xy}\) =const at t\textgreater{}0), it propagates inside towards the opposite surface as a wave traveling with sound speed. It reflects from that opposite side with negative amplitude (that surface is fixed and it’s movement is not allowed). The reflected wave interferes with the incoming wave forming a complex pattern of pressure inside of sample volume. A situation like that has a well known analytical solutions in case of a medium that is characterized by linear response. For instance, time dependence of amplitude, $a(t)$ of sound wave is a solution of second-order differential equation and may be written as \cite{Hayek}, \cite{Thornton}:

\begin{equation}
a(t) =1 - \mathrm {e}^{-\zeta \omega t} \sin \left(\sqrt {1-\zeta ^2} \omega t+ \varphi \right)/\sin(\varphi),
\label{eq:damp}\end{equation}

with phase \(\varphi\) given by \(\cos \varphi =\zeta\) and \(\omega=(k/m)^{1/2}\) is called the \textit{undamped angular frequency}, \(\zeta\) is called the \textit{damping ratio} and is related to energy losses. Quantities $k$ and $m$ in a simple classical model are spring constant and oscillating mass. 

In case of a linear in time ramp pressure applied, $P^0_{xy}(t)=h\cdot t$, where h is the pressure rate change in time, response of a linear medium is known analytically and it is given by:

\begin{equation}
P_{xy}/h = t -\tau \cdot (1-e^{-t/\tau}), 
\label{ramp00}
\end{equation}

where $P_{xy}$ is component of pressure tensor as reported by LAMMPS.

Equations like these above, \ref{eq:damp} and \ref{ramp00}, are often found in textbooks on automation and they describe a broad range of phenomena, mostly in engineering. 

The above equations are written for xy component but they are more general, though parameters such as $\tau$ above may depend on components, due to material anisotropy and sample geometry. Hence, in case of linear ramp force applied there is a time lag of pressure at the surface that additionally decays exponentially with time. Equation \ref{ramp00} is  however a solution of first order differential problem, with second order contributions (leading to oscillations) omitted.

We have performed a series of simulations for the applied pressure rate changes $h$ from between 1 MPa/ps to 200 MPa/ps and found that surface pressure as reported by LAMMPS is well described by the function:

\begin{equation}
P_{xy} = h \cdot t - A \cdot h \cdot \sin (\omega t)
\label{ramp01}
\end{equation}

For sample of 100 {\AA} size in Y-direction the period of oscillations is around 10.7 ps, i.e. that $\omega$ is around $0.6 ps^{-1}$, while value of A is about 0.1 ps, when time step used in MD simulations is 0.001 ps. Similar oscillations are notoriously found in MD studies \cite{Monnet}.

These results show us that in case of a ramp pressure applied, $P_{xy}(t)$ is close enough (in a reasonable experimentally sense) to applied $P_0(t)$ for very low values of $h$ only, of around just 1 MPa/ps or less, which is hardly computationally convenient for MD simulations that usually require a lot of testing yet in preliminary phase of any studies. On another hand ramp rates of the order of 100 MPa/ps are too fast for achieving quasi-stable physical conditions (dynamical stresses) inside the sample volume.

In this work we concentrate on using the first method of modeling. i.e. that one when pressure is applied abruptly at t=0. Our aim is gaining a deeper understanding of stress dynamics and timescales involved for a proper design of the simulation setup and later analysis of results.

\section{Atomic configuration}

\subsection{Creating oriented Steel 310S with dislocations.}
\label{creating-an-fcc-sample-of-iron-fe}

For creation of samples of steel 310S (FCC structure) we used either LAMMPS \cite{LAMMPS} or Atomsk\footnote{https://atomsk.univ-lille.fr} \cite{Atomsk}, as well custom-written Perl\footnote{https://www.perl.org/} scripts. The weight contribution of Fe-Ni-Cr atoms is 0.55-0.20-0.25, as in specifications\footnote{https://www.azom.com/article.aspx?ArticleID=4392}. After testing available interatomic potentials (\cite{Bonny}, \cite{Artur},\cite{Mendelev},\cite{ZhouXW},\cite{Wu}) we choose the EAM potential of Bonny et al. \cite{Bonny} as the one that reproduces correctly basic physical properties of the material and as the most efficient one computationally. Moreover, that potential has been developed to model defects in steel of similar composition and therefore is suitable for our planned future modeling. Some results were computed by using Artur potential \cite{Artur} as it was observed that Bonny potential in some situations leads to computational instabilities of unclear for us so far origin.

\begin{figure}[ht]
\centering
\includegraphics[scale=0.9]{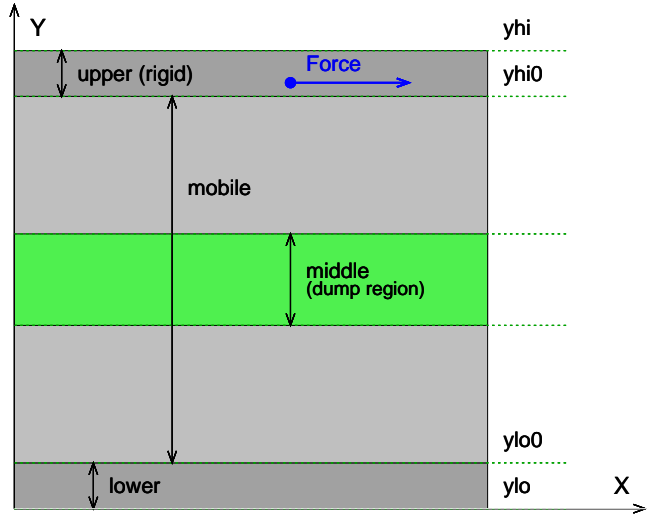}
\caption{Sample crossection in X-Y directions for modeling dynamics of dislocations (not to scale). For a proper capture of dislocation dynamics we need a long size in X-direction, which is direction of dislocation movement (and Burgers vector orientation). X-Y plane is the gliding plane while dislocation line is perpendicular to it, i.e. it is in Z-direction.
}\label{fig:sample}
\end{figure}

Samples have been imposed to potential energy minimization, checked for density and elastic constants at low temperatures and at room temperature (most of results we report on here were computed at T=50K). 

It is important to include details of samples geometry and computational methodology since, as it will become evident, these have profound impact on results obtained and on their interpretation.

We use a typical \cite{Osetsky} simulation setup applied in similar modeling, as shown in Fig. \ref{fig:sample}.

For sample sizes and positions of dislocations we use the following convention. Sizes in Y-direction are multiplicity of FCC unit cell in that direction. The sample is oriented, with X=$[1\bar{1}0]$ (which is direction of the Burgers vector \textbf{b}), Y={[}111{]} (which is direction normal to the glide plane), and Z=$[\bar{1}\bar{1}2]$). That unit cell in Y-direction has size of 6.16 {\AA} and it consists of atoms of hexagonal structure arranged in 3 layers in X-Z plane separated by 2.06 {\AA} in Y-direction. Names of samples depend on sizes in Y-direction. \textbf{M+N} means that the position of dislocation is \textbf{M} unit cells from sample bottom (i.e. where Y=0) followed by \textbf{N} unit cells to the surface where pressure \(P^0_{xy}\) is applied. Hence, samples used here for dislocation dynamics studies have the geometries 2+2, 4+4, 8+8, 16+16. Additionally, there is a series of calculations on samples 8+X, where \(X=n \cdot 8\) and n ranges from 1 to 7, and a sample named 48 (total size in Y direction) which is in 3 variants: 24+24, 16+32 and 32+16. 

The shear surface stress $P^0_{xy}$ is created by applying a force to atoms in the \textbf{upper} region while on the bottom surface of the \textbf{lower} region conditions are imposed of zero forces and velocities of atoms. The force \textbf{F} applied to atoms must be oriented in X-direction and have a value such that when summed for  all atoms in the \textbf{upper} region and divided by the surface area in X-Z direction it will result in desired value of pressure $P^0_{xy}$.

We find that the width of \textbf{upper} and \textbf{lower} regions in Fig. \ref{fig:sample} may be made as small as the distance between atomic layers in Y-direction, i.e. that these regions contain one only layer of atoms.

Temperature is stabilized with NVT thermostating in \textbf{upper} and \textbf{lower} sample regions only, while the interior of the sample remains under quasi-adiabatic conditions, with some only heat exchange with these two temperature-controlled regions. Average sample temperature of entire sample volume remains constant, with RMS fluctuations not exceeding 0.7K, for any sample size, at T=50K, provided that no dislocations are there and applied pressure is below a few GPa. Otherwise strong temperature instabilities may occur. External pressure is transferred to sample interior through NVE integration updating atoms positions and velocities.

For saving on disk storage space and data analysis time it is sometime convenient to dump data from a narrow region only, named \textbf{middle}, which encompasses the core of dislocation.

As a time step in MD simulations of dislocations movement we used mostly 0.001 ps, a time which is the largest possible to speed up the computation and is still not disturbing the accuracy of modeling. In some cases, as explained, other time step was used.

Dislocations of edge type (lines) were created with the help of Atomsk or by displacing atoms with the method of Carpio et al. \cite{Carpio} implemented ourselves \footnote{Carpio et al. \cite{Carpio} do not explain explicitly that there are three ranges of x-y values where their equations need slight modifications; We are willing to provide additional explanations on a request, including a sample code for generating displacements field.}. An overview of the methods of dislocations creation is found, e.g., in \cite{zhang}. When using Atoms, a perfect $\frac{1}{2}[110]$ dislocation with a stacking fault is obtained by joining two crystals in Y direction that have length change by $\pm a/2$ in oriented X direction, where $a$ is lattice constant in that direction. That dislocation splits into two Shockley partials with Burgers vectors $\frac{1}{6}[\bar{2}11]$ and $\frac{1}{6}[\bar{1}2\bar{1}]$ \cite{Nabarro} repelling each other and being attracted by stacking fault between them. In result a stable equilibrium position is reached when they are separated for about 66 {\AA} in X-direction. Under stress they both move the same way, with some only fluctuations of the distance between them.

Dislocation visualization was done with Ovito \cite{Ovito}, while for extraction of dislocation data Ovito were used, Ovitos (Python extension libraries to Ovito), Crystal Analysis Tool\footnote{https://gitlab.com/stuko/crystal-analysis-tool} \cite{CAT} and/or our own home-brewed scripts in Perl.

Statistical averaging of physical quantities along X direction was done in Perl as well, with resolution of 2.06 {\AA} in Y direction, which is the distance between neighboring crystallographic planes in that direction. Data analysis was often accompanied by creating movies with process automation utilizing bash, Perl and gnuplot\footnote{http://www.gnuplot.info} scripts, as well {ffmpeg}\footnote{https://www.ffmpeg.org/}.

\begin{figure}[ht]
	\centering
	\noindent\includegraphics[scale=1.0]{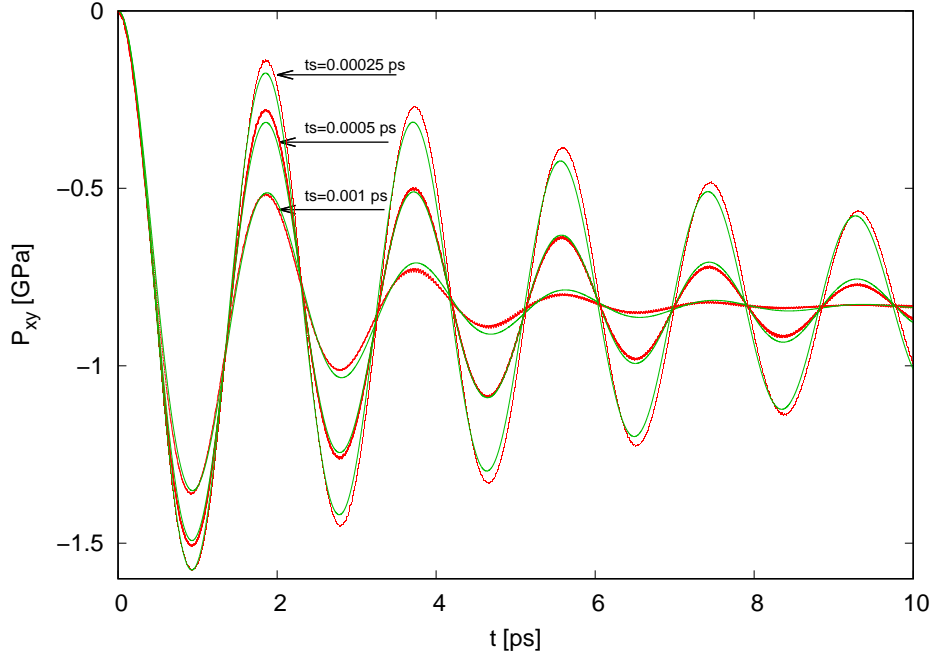}
	\caption{Shear pressure $P_{xy}$ as a function of time reported by LAMMPS at the upper surface of sample of size Y=4 (4 unit cells). Three pairs of curves are shown for 3 values of time-step used in LAMMPS, as shown in the figure. Red data points are results of MD simulation while green curves are computed with Eq. \ref{eq:dampingP}. The applied pressure at the surface, $P^0_{xy}$, is 1 GPa.}
	\label{steps015}
\end{figure}

\begin{figure}[ht]
	\centering
	\noindent\includegraphics[scale=1.0]{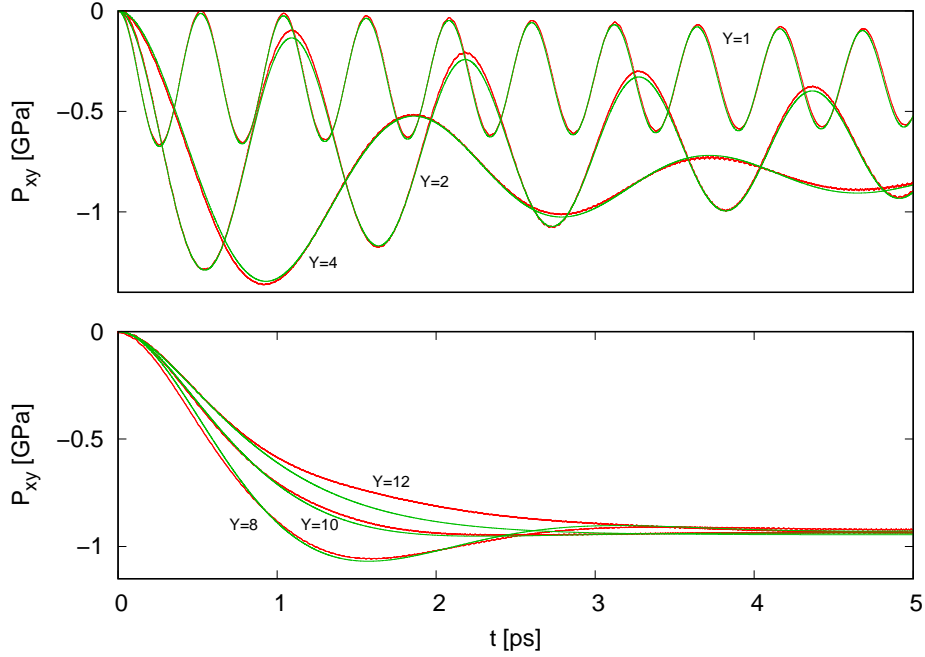}
	\caption{$P_{xy}(t)$ dependencies, similar to these in Fig. \ref{steps015}, for samples of sizes from Y=1 to Y=12, when the same time-step is used in MD simulations, $ts=0.001$ ps. Red data points are results of MD simulation while green curves are computed with Eq. \ref{eq:dampingP} or \ref{eq:dampingPP}. The applied pressure at the surface, $P^0_{xy}$, is 1 GPa. In case of curves at the bottom figure the damping parameter $\zeta$ approaches (Y=8) and exceeds the critical damping ratio ($\zeta=1$). The curve for Y=12 is in overdamped regime. In this case we see that green lines, as computed by Eq. \ref{eq:dampingPP} start to depart from MD simulation results. That difference increases when $\zeta$ grows more above 1.}
	\label{steps016}
\end{figure}

\subsection{Pressure response of sample.}
\label{Stress distribution}

The simplest way of studying stress dynamics is the measurement of stress at the sample surface after external pressure is applied, as shown in Fig. \ref{steps015}. Strong oscillations of $P_{xy}(t)$ are found if proper simulation conditions are used. We observe that these curves are excellently well described by the dumped harmonic oscillator response to a step time pressure change, as given by Eq. \ref{eq:damp}. In this case pressure $P_{xy}$ has a negative sign as a response to positive pressure applied at the surface, $P^0_{xy}$. Hence, to fulfill the boundary conditions let us rewrite Eq. \ref{eq:damp} in the following form:

\begin{equation}
P_{xy}(t) =A\cdot \left[\mathrm {e}^{-\zeta \omega t} \sin \left(\sqrt {1-\zeta ^2} \omega t+ \varphi \right)/\sin(\varphi)-1\right],
\label{eq:dampingP}\end{equation}

Parameters $\omega$ and $\zeta$ are material specific. The damping parameter $\zeta$ depends also on details of the MD simulation process (in particular, it is proportional to time-step). Angular frequency $\omega$ depends also on sample size in Y-direction. The all three green curves in Fig. \ref{steps015} have been computed by using the same values of $A$ and $\omega$, and value of $\zeta$ scaled proportionally to time-step $ts$. 

Parameter $A$ in Eq. \ref{eq:dampingP} has a dimension of pressure and it scales up linearly with applied pressure $P^0_{xy}$ (at least for $P^0_{xy}$ not exceeding a few GPa), but its value is lower than $P^0_{xy}$. It approaches $P^0_{xy}$ only for $\zeta \gg 1$ (which becomes large when sample size in Y-direction is large). This is seen better in Fig. 
\ref{steps016}, where $P_{xy}(t)$ dependencies are shown for samples of different sizes, from Y=1 to Y=12. 

The damping parameter $\zeta$ in Eq. \ref{eq:dampingP}
must be lower than 1, since we have a $\sqrt {1-\zeta ^2}$ factor there. In case of $\zeta > 1$ the above equation is replaced by a one usually written  as a sum of two exponential contributions. By using relations: $sinh(x)=-isin(ix)$ and  $arcosh(x)=ln(x+\sqrt{x^2-1})$, we may formulate Eq. \ref{eq:dampingP} for $\zeta > 1$ in a similar form:

\begin{equation}
P_{xy}(t) =A\cdot \left[\mathrm {e}^{-\zeta \omega t} \sinh \left(\sqrt {\zeta^2-1} \omega t+ \varphi \right)/\sinh(\varphi)-1\right],
\label{eq:dampingPP}\end{equation}

As seen however in Fig. \ref{steps016}, for large samples (for $\zeta>1$) the above equation does not reproduce exactly the results  computed in MD simulations.

Equations \ref{eq:damp}, \ref{eq:dampingP} and \ref{eq:dampingPP} are obtained as solutions of the second order differential equation. Therefore there must be two parameters for satisfying general boundary conditions. One of them is amplitude $A$ of obvious physical meaning. The second one is related to the initial velocity of wave propagation to the medium at the surface. That is why parameter $\zeta$ is found to depend on the time-step in MD simulations.

\begin{figure}[ht]
	\centering
	\noindent\includegraphics[scale=0.9]{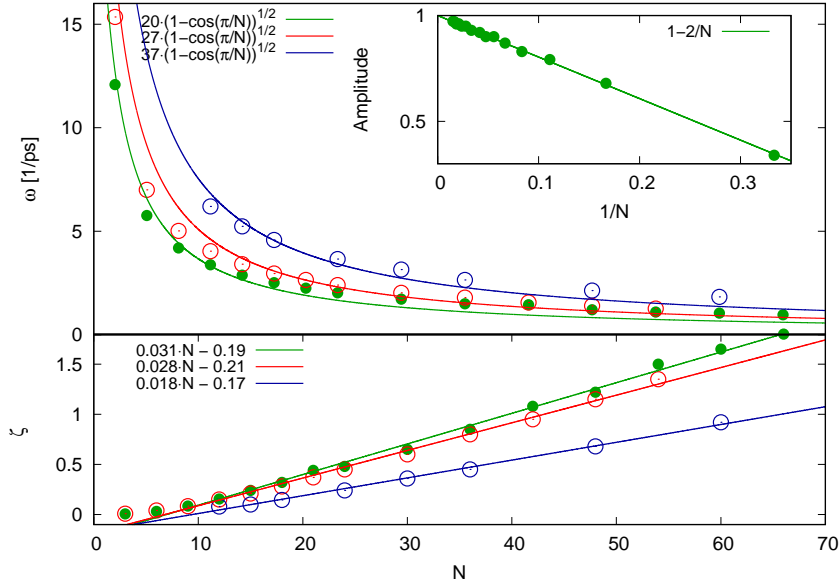}
	\caption{Dependencies of $\omega$, $\zeta$ and amplitude of oscillations $A$ used in equations \ref{eq:dampingP} and \ref{eq:dampingPP} on number of atomic layers in Y-direction, N, as deduced from analysis of MD simulations. The data points in green (full circles and curves) are obtained for simulations using Artur potential, while these in red and blue for Bonny potential. Data points and curves in blue are for direction $[11\bar{2}]$ while all other for Y in direction $[111]$.}
	\label{omega003}
\end{figure}

Detailed dependencies of $\omega$, $\zeta$ and amplitude of oscillations $A$ on number of atomic layers in Y-direction is shown in Fig. \ref{omega003}. The number N is simply multiplicity by 3 of the number of unit cells of [111] oriented FCC crystal.

Amplitude of oscillations does not depend on potential used or on crystal orientation and it has 1/N relationship, as shown in Fig. \ref{omega003}. Damping coefficient, shown there, is for MD time-step of 0.001 ps and it retains linear dependence on N, regardless of potential used or crystal orientation. It is slightly only different for potentials of Artur and Bonny, and it depends on crystal orientation. The largest change is found in $\omega$: larger values are found for Bonny potential, as expected, since that potential produces a larger values of elastic parameters.  Angular frequency $\omega$ depends on crystal orientation as well.

Since amplitude of oscillations is inversely proportional to N, it reaches 1 only in the limit of infinite size of sample, while for small sample sizes it is significantly lower than one. That means that volume average pressure value inside of a small sample is always lower than the pressure applied at sample surface. 


Dependence of angular frequency on N, $\omega(N)$, may be interpreted within a simple classical model 
\cite{Thornton} in which all layers of atoms separated by about 2.06 {\AA} in Y-direction are treated as masses connected by springs. This approach was found useful in case 
of interpreting, for instance, Raman spectra in a few layers graphene \cite{Lui}, \cite{Tan}, \cite{Koziol}. We find that a solution to this mathematical problem proposed by Tan et al. \cite{Tan} may be used as a description of the observed $\omega(N)$ dependence. In \cite{Tan} they write angular frequencies of the possible set of solutions for N layers in the following form, for the $i$th vibrational mode:

\begin{equation}
\omega^2_i = \omega^2_0 \cdot \left( 
1-cos\left( \frac{(i-1)\pi}{N} \right)  
\right),
\label{eq:omega}\end{equation}

where $i=2 \dots N$, and $\omega^2_0$ is the angular frequency of the classical harmonic oscillator.

There is a significant difference however between their case and ours. In \cite{Tan} the highest possible frequency mode is used (when $i=N$), as the one active in Raman spectroscopy, while we find the lowest possible frequency (when $i=2$) as vibrational mode suitable here. Therefore we may write:

\begin{equation}
\omega = \omega_0 \cdot \sqrt{ 
1-cos\left( \frac{\pi}{N} \right)  
},
\label{eq:omega2}\end{equation}

As shown by the solid line in the upper part of Fig. \ref{omega003}, the angular frequency of damped oscillations are well fitted by Eq. \ref{eq:omega2} when value of $\omega_0$ of about $27/ps$ is used.

Another important difference between results of Tan et al. and that of ours follows from different boundary conditions: in our case lowest amplitude of atoms oscillations is observed at surfaces and maximum is located at sample geometrical center.

For estimating the value of $\omega_0$ in a model of harmonic oscillator we need to know $k$ parameter in interatomic potential dependence on distance between atoms, when an atom embedded in a crystal lattice moving as a whole probes the potential well at various positions: $U=kx^2/2$, where $x$ is distance from the minimum of potential position. In order to find out this, a LAMMPS script was created with a frozen (not moving) FCC lattice and a set of $10^4$ random potential values for iron atom  were computed (for random positions within a distance of 0.2 {\AA} from potential minimum). 

It was found that $k$ is about 8 eV/{\AA}$^2$ (potential is in fact not exactly harmonic and it is not even symmetric with respect to zero).

With Fe atomic mass of 55.845 we obtain  $\omega_0=\sqrt{k/m}=37.0/ps$ or in terms of frequency and energy it is 5.9 THz, or it is 24 meV. That value is in a reasonable agreement with computed \cite{Muller} or experimentally determined phonon energies in Fe \cite{Zarestky}, \cite{Stankov} or steel \cite{Danilkin}, \cite{Urbassek}.

\section{Dynamics of internal virial stress.}
\label{dynamics-of-stress-penetration}

Pressure $P_{xy}$ discussed so far is a physical quantity which is observed at the surface of sample and it's value is a result of summation of contributions to stress within entire sample interior. Dynamics of dislocations however depends mostly by stress distribution at dislocation core and around it. Therefore what we need to know for studying it is the internal stress distribution, which additionally is of changing in time nature. Introducing dislocations into sample volume changes distribution of stress there and changes also penetration dynamics of externally applied pressure. We can not assume that contributions to local stresses are the simple sum of stress field caused by dislocation and the one due to penetrating the material interior external wave of pressure, since the stress at the core of dislocation is beyond the linearity limit. At the same time stress field changes due to dislocation movement. 

Throughout this article we analyze the virial contribution from interatomic potential as given by the second term in Eq. \ref{eq:virial}, since in case of EAM potential of Bonny et al. \cite{Bonny} no other terms there are possible, while the kinetic energy term contribution to stress is of the order of $\sim 10^{-3}$ only of that of potential energy, at temperature T=50K, and therefore it can be ignored.

The LAMMPS implementation of virial stress by Thompson et al. \cite{Thomson}, of a microscopic quantity that may be interpreted as a local, atomistic stress tensor (Eq. \ref{eq:virial}), passes the basic test of validity, in our case for steel 310S, which is the condition that the sums of all diagonal virial stress tensor components for all atoms in sample volume and at any time must be equal to total pressure with minus sign:

\begin{equation}
\left(S_{xx}+S_{yy}+S_{zz}\right)/3V = -P
\label{eq:virial2}\end{equation}

We find that a similar relation is observed for the surface pressure shear component $P_{xy}$ as well, i.e.: 

\begin{equation}
S_{xy}/V = -P_{xy}. 
\label{eq:virial3}\end{equation}

In Eq. \ref{eq:virial3} $S_{xy}$ is value of virial stress tensor component $xy$ averaged over all sample atoms.

One needs to take care of the conversion between $S_{ij}$ and $P_{ij}$ to avoid confusion: virial stress, as reported by LAMMPS, must be divided by an average volume of atoms (in units of {\AA}$^3$, when \textit{metal} units are used), and hence it is material specific, while it is reported in LAMMPS in units of pressure. In our case, for Steel 310S that conversion factor is 11.21, i.e. $S_{xy}[GPa] = -11.21\cdot P_{xy}[GPa]$.

\begin{figure}[ht]
	\centering
	\noindent\includegraphics[scale=0.9]{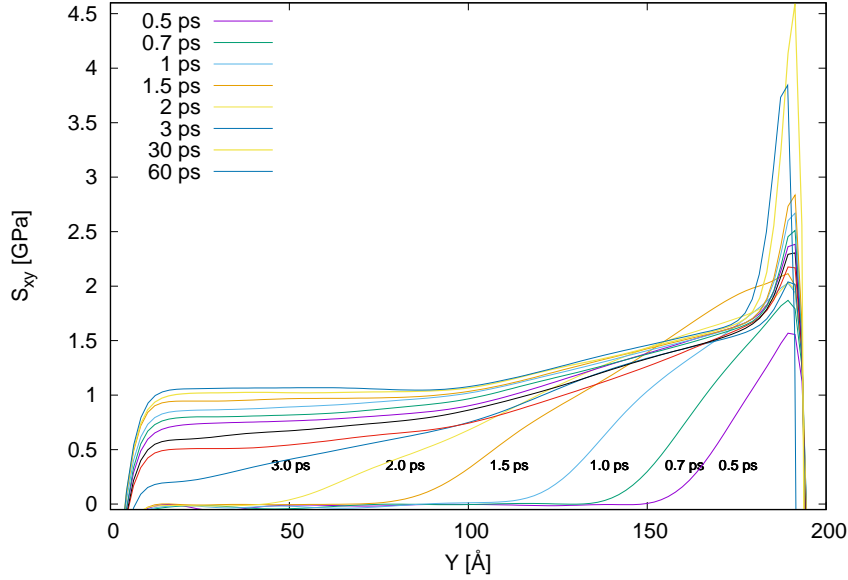}
	\caption{At T=50K pressure \(P_{xy}\) = 1.8 GPa is applied at t=0 at the upper surface located at around Y=200 {\AA}. Dislocations are located in geometric Y-center at around 100 {\AA}. The lines show \(S_{xy}\) profiles at time moments as described in the Figure. At around 30 ps the profile below sample center saturates and becomes stable while above dislocation lines it still evolves with time. }\label{fig:slices380}\end{figure}

The relation expressed by Eq. \ref{eq:virial3} is fulfilled with high accuracy during all dynamic simulations and it can be used as a test of validity of any computation: any deviations from it must be treated as an almost certain signature of computational instabilities or methodological errors in the modeling process itself.

Figure \ref{fig:slices380} shows typical results on virial stress distribution $S_{xy}$ in a 16+16 sample containing a dislocation line in its geometrical center in Y-direction. 

\begin{figure}[ht]
	\centering
	\noindent\includegraphics[scale=0.8]{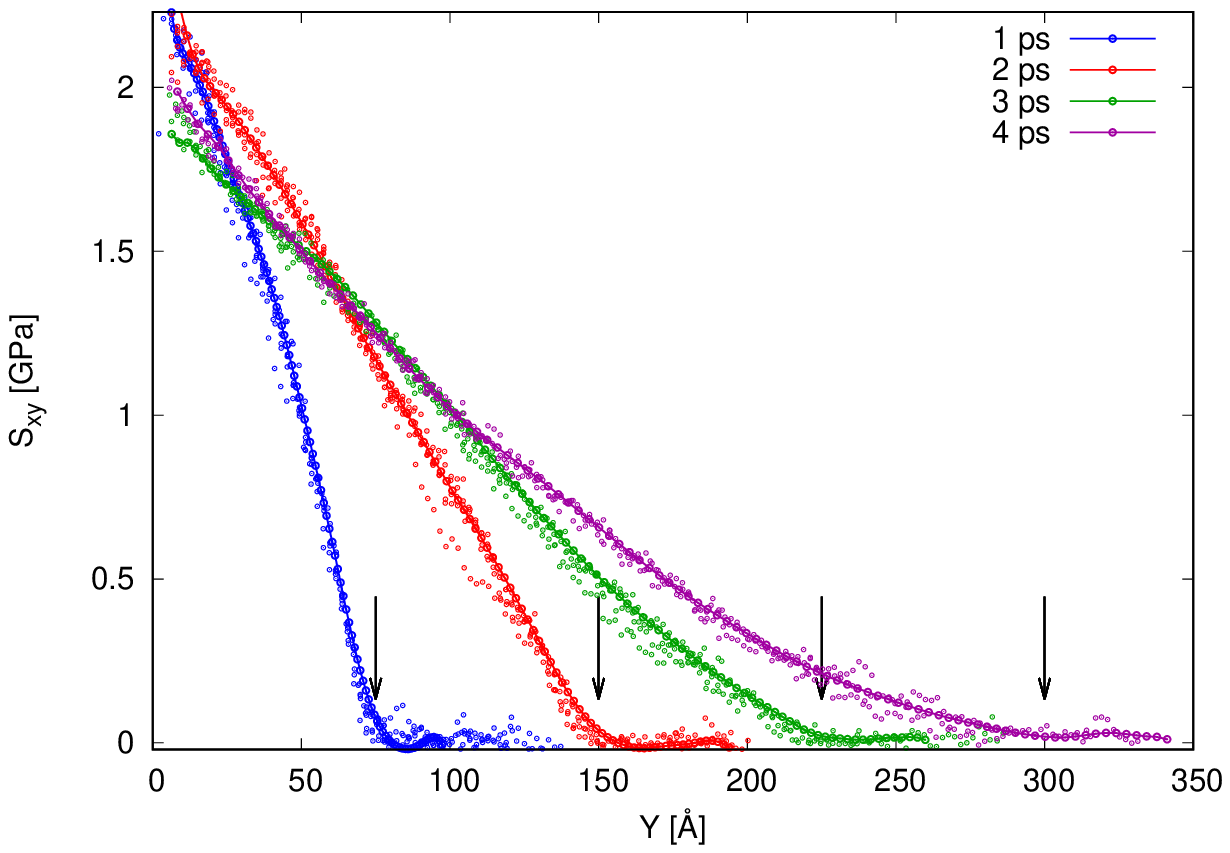}
	\caption{Large number of curves of initial \(S_{xy}\) profiles are used for estimating the speed of pressure propagation into the medium. Pressure $P^0_{xy}$=2 GPa is applied at t=0 to surface of samples at Y=0. Arrows indicate approximate position of the wave front for each data set, corresponding to the speed of propagation of 75 {\AA}/ps.}\label{fig:speed}\end{figure}

The profiles of \(S_{xy}\) averaged over X-direction (which is direction of dislocation movement, i.e. it is perpendicular to dislocation alignment along Z-axis) are shown in Fig. \ref{fig:slices380} as the pressure penetrates successively inside of the sample interior. Values of \(S_{xy}\) saturate for times larger than about 30 ps for values of Y-coordinate that are below the dislocation line position (which in this case is at around Y=100 {\AA}). A large peak of \(S_{xy}\) near the surface initially increases with time and starts to decrease slowly at around the same time when saturation of \(S_{xy}\) is observed below the dislocation line position.

Results of Fig. \ref{fig:slices380} clearly indicate the existence of a profile of pressure spreading into the interior of sample. Therefore we performed calculations for samples of various sizes in Y-direction in order to determine accurately the speed of pressure wave penetration. Similar analysis must always be done on a large number of data, as fluctuations of internal stresses are large: Probability Distribution Function for $S_{xy}$, as computed by us for all atoms, resembles a Gaussian function and at T=50K and at zero applied pressure has a half-width of about 3.0 GPa. 

Results of Fig. \ref{fig:speed} indicate that the penetration depth, as marked approximately by arrows, is linear with time and the penetration speed is about 75 {\AA}/ps, which is close to the speed of sound waves, as estimated by us.

\begin{figure}[ht]
	\centering
	\noindent\includegraphics[scale=0.9]{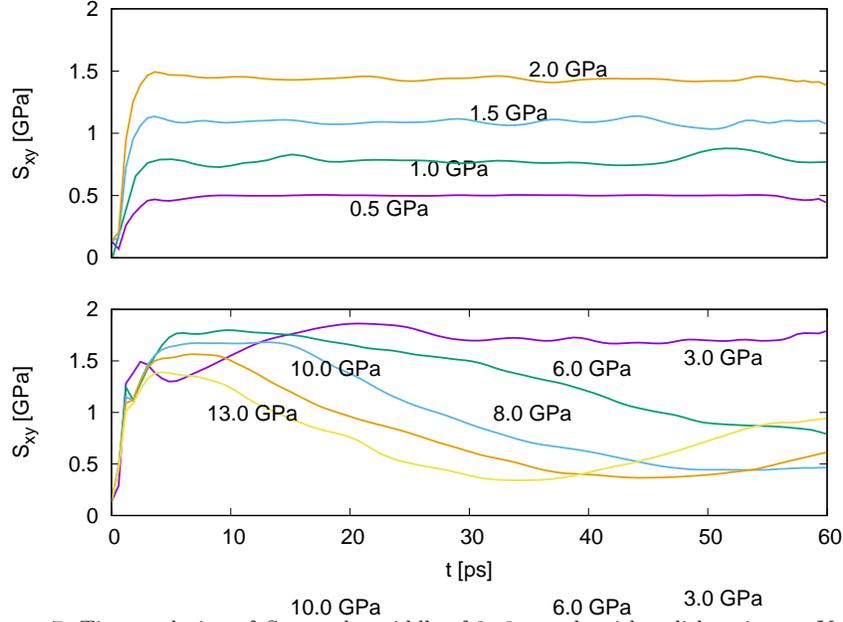}
	\caption{Time evolution of \(S_{xy}\) at the middle of 8+8 sample with a dislocation, at Y-value that is the dislocations location, i.e. at about Y=50 {\AA}). The upper part shows the evolution at lower pressures, the lower part at higher values of \(P^0_{xy}\), as labels indicate. It is remarkable that at high values of pressure applied \(S_{xy}\) changes dramatically with time and its values are much lower than \(P^0_{xy}\). }\label{fig:low-high}\end{figure}

More understanding of dynamics of stress penetration comes from analysis of Fig. \ref{fig:low-high} where a few examples of time evolution of \(S_{xy}\) at the middle of 8+8 sample (at the dislocations core location) are shown. After an initial fast increase of the stress, a quasi-stable \(S_{xy}\) values are observed when the pressure applied is relatively low (top curves). At higher pressures however (lower part of Fig. \ref{fig:low-high}) large dynamic changes are found.

At applied pressures \(P^0_{xy}\) above around 2.4 GPa time dependence of \(S_{xy}\) becomes increasingly unstable, with a long period damping oscillations dominating. This is found when samples with dislocations are investigated. 

\begin{figure}[ht]
	\centering
	\noindent\includegraphics[scale=0.9]{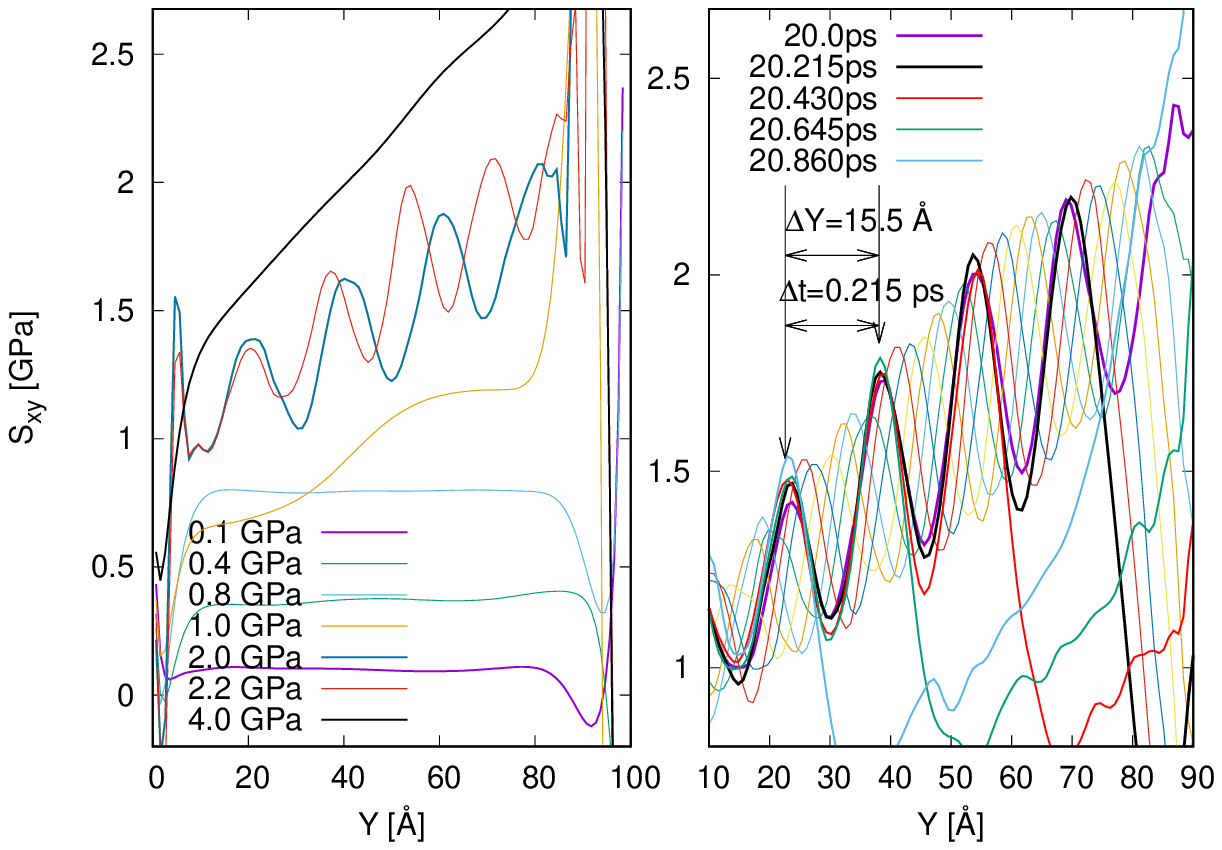}
	\caption{On left: Profiles of \(S_{xy}\) across 8+8 sample with no dislocations at 50 ps after the external pressure \(P_{xy}\) has been applied to the surface at Y of around 100 {\AA}, for several values of \(P^0_{xy}\).
		On right: Spatio-temporal oscillations of $S_{xy}$ in a 8+8 sample without dislocations at applied surface pressure $P^0_{xy}=2.2 GPa$. Waves of pressure pushed into sample interior have a well defined wavelength and oscillation period.}
	\label{fig:profiles0}
\end{figure}

It is interesting to see how the stress penetrates in case of samples without dislocations. This is shown in Fig. \ref{fig:profiles0} (left), where profiles of \(S_{xy}\) across 8+8 sample with no dislocations, at time 50 ps after the external pressure \(P^0_{xy}\) has been applied to the sample surface located at Y of around 100 {\AA}. At certain range of applied pressure (2.0 to 2.5 GPa in that Figure) well observed spacial oscillations of \(S_{xy}\) are found. At higher \(P^0_{xy}\) values the stress profile become nearly linear with y-coordinate.

These results strongly suggest the existence of spatial oscillation in sample volumes at certain values of applied pressure. Therefore it was natural to search for time oscillations as well. First we found time-dependent oscillations (a traveling wave) for a sample 16+16, with time resolution of 0.1 ps and we were able to determine wavelength of oscillations to be about 14 {\AA} while their period of about 0.2 ps. These values correspond to wave propagation speed of 70 {\AA}/ps. 

After repeating computation on a sample of 8+8, with a higher time resolution of 0.005 ps, as shown in Fig. \ref{fig:profiles0} (right), we find with a better accuracy the wave propagation speed of 72 {\AA}/ps and notice also that this speed does not depend on sample size in Y-direction. That is shown in Fig. \ref{fig:profiles0} (right), as pressure penetrates to the sample interior from the surface located at about Y=100 {\AA}. A few waves that coincide are plotted with thicker pen. We can determine  well the period of oscillations and distance between wave maxima, and find that $\omega=(2\pi/\Delta t)/ps= 29.2/ps$. Hence, the observed oscillation frequency is in a good agreement with that one expected, based on classical oscillator equation, and in agreement with angular frequencies found out from measured $\omega(N)$, as shown in Fig. \ref{omega003}.

\section{Velocity of dislocation movement.}
\label{velocity-of-dislocation-movement}

Once we have established the basic understanding of \(S_{xy}\) field penetration dynamics, we may start analysis of dislocation displacements.

\begin{figure}[ht]
	\centering
	\noindent\includegraphics[scale=0.9]{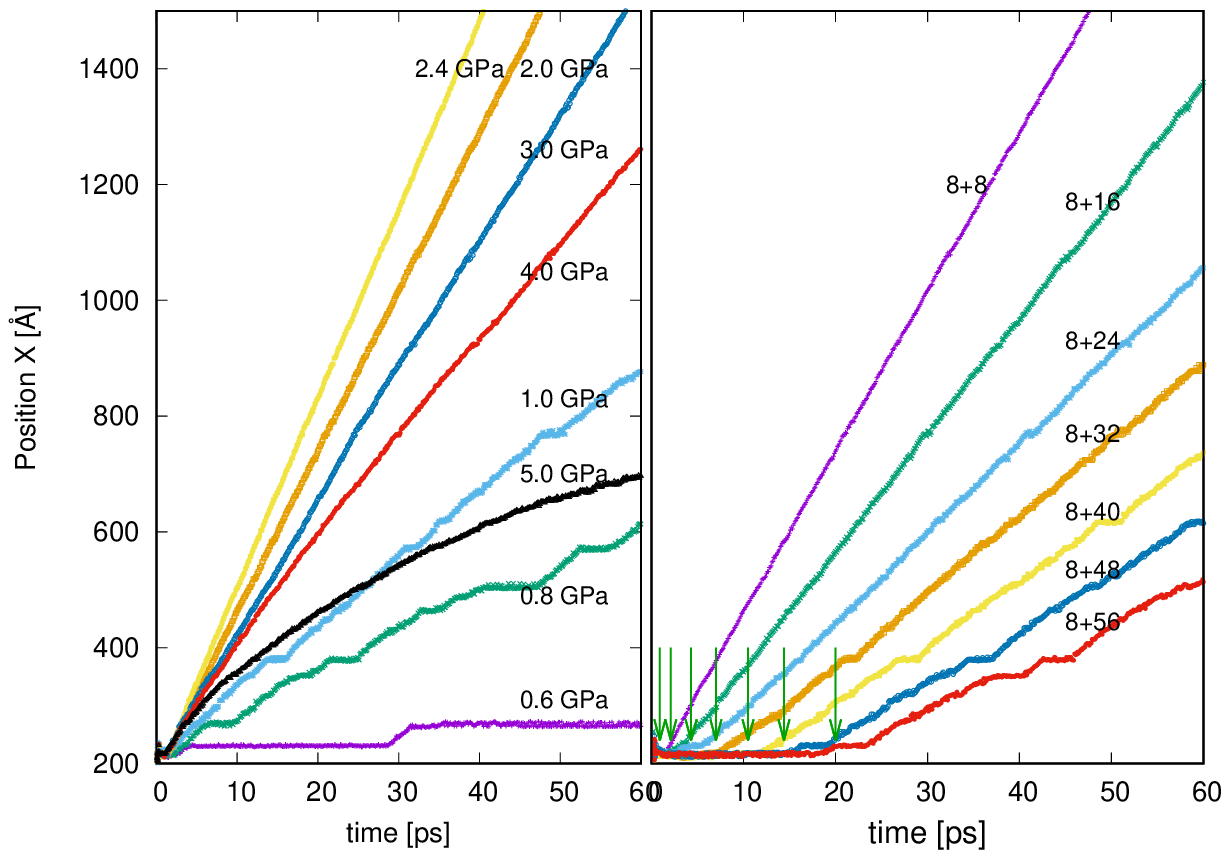}
	\caption{Typical dependencies of dislocation displacements as a function of time after the pressure \(P^0_{xy}\) is applied at one surface of the sample. On the left for several values of the applied pressure $P^0_{xy}$. On the right
as function of time after pressure \(P^0_{xy}\) of 2 GPa is applied, for a series of samples of different sizes in Y-direction. Dislocations are located 8 unit cells from bottom surface for every sample. The green arrows on time-axis indicate approximate moments when dislocation starts to move.}\label{fig:displace0}\end{figure}

Figure \ref{fig:displace0} on the left shows typical dependencies of dislocation displacement as a function of time after the pressure \(P^0_{xy}\) is applied for pressure values as indicated there. For the most bottom data points at pressure of 0.6 GPa  dislocations move at some random moments only and it is ambiguous determining its velocity. The data points at 0.8 G Pa move almost all the time. At the highest speeds (2.4 GPa in this case) displacements are monotonic in time and form nearly ideal straight lines. At higher pressures however there is an abrupt change to curves with a strong curvature, and again there is some ambiguity in determining the speed of dislocation movement (curve for 5.0 GPa). We used for that the slope of curves on the side of large time. At yet higher pressures again an irregular movement is found and dislocations gradually disintegrate.

Hence, we observe that there exists a maximum of dislocation velocity at pressures around 2.4 GPa in case of the sample 8+8 of Fig. \ref{fig:displace0}. A question arises if sample geometry has any influence on computational results. To check that modeling was performed on a series of samples named 8+X, where \(X=n \cdot 8\) and n ranges from 1 to 7, as shown in Fig. \ref{fig:displace0} on the right. In these samples dislocation core is located 8 lengths of unit cell from sample bottom while pressure is applied at the top surface which is $X=n \cdot 8$ lengths of unit cell from dislocation core.

We find that distance of surface from the dislocation line not only influence dislocation velocity but also the time when it starts to move.

\begin{figure}[ht]
\centering
\noindent\includegraphics[scale=0.8]{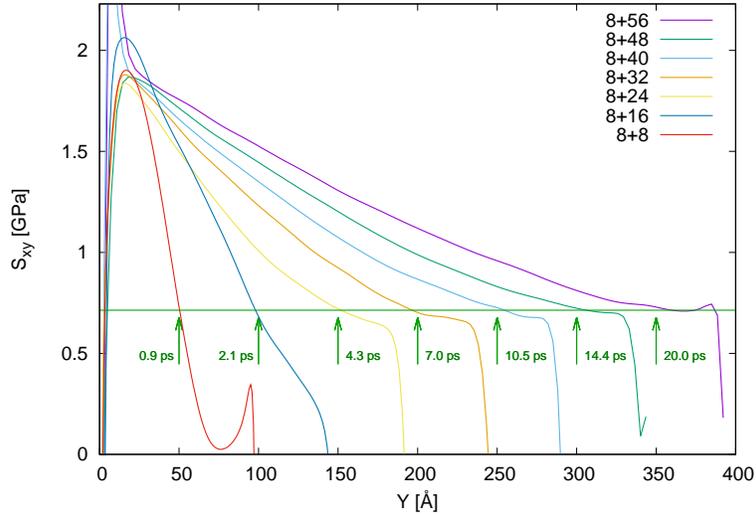}
\caption{Combined profiles of \(S_{xy}\) for the same samples as in the right of Fig. \ref{fig:displace0}, at time moments when dislocations start to move, indicated on both figures by green arrows. It is found that \(S_{xy}\) profiles cross at these moments the green horizontal line of \(S_{xy}\)=0.75 GPa. Curves are smoothed for clarity with Bézier function.}\label{fig:8GPa}\end{figure}

Figure \ref{fig:8GPa} shows \(S_{xy}\) profiles for the same samples as in Fig. \ref{fig:displace0}, at time moments indicated there by green arrows, when dislocations start to move. Each of these profiles intersects with a straight horizontal line at \(S_{xy}\) = 0.75 GPa, indicating that this is the critical value of stress needed for dislocation movement. Moreover, position of these points on Y-axis is the same as position of the core of dislocations.

\subsection {Velocity scaling with pressure and sample sizes.}

\begin{figure}[ht]
\centering
\noindent\includegraphics[scale=0.9]{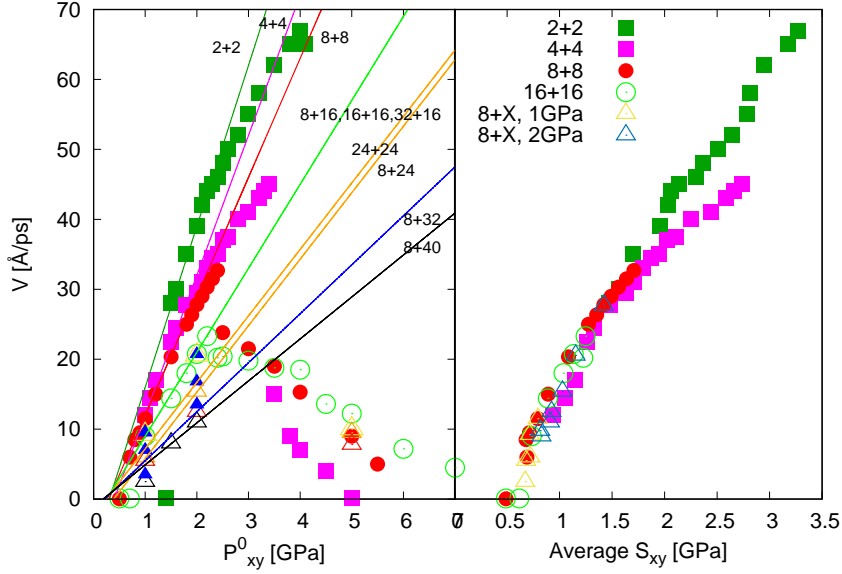}
\caption{On the left: Dislocations velocity as a function of applied pressure \(P^0_{xy}\) for samples of various sizes in Y-direction and various distances between the surface where pressure is applied and dislocations positions.
	On the right: Dependence of dislocations velocity in \textbf{region I} and \textbf{region II} is found to scale well with the average stress \(S_{xy}\) in the sample space region between the sample bottom and location of dislocations.
	Legend in the right part of figure refers to data points in the left side as well.
}\label{fig:Vscaling}\end{figure}

Figure \ref{fig:Vscaling} on the left shows a collection of velocity data points from computations performed on samples of different sizes and pressures \(P^0_{xy}\) as well various distances between the surface where pressure is applied and dislocations positions. Clearly three ranges of applied \(P^0_{xy}\) can be distinguished (these ranges change with sample geometry). 

\textbf{I region}: At lowest pressure below around 1 GPa, velocity is small (and determined with large uncertainty) due to non-monotonic dislocation movement, marked with periods of no movement and abrupt jumps in speed. The value of 1 GPa coincides well with the value of internal stress of around 0.75 GPa, needed for displacing dislocations, as deduced from Fig. \ref{fig:8GPa}.

\textbf{II region} is between around 1 GPa and 2.4 GPa or 2.2 GPa for samples 8+8 and 16+16, respectively. Dislocation movement in this region is monotonic in time, and displacement is characterized by a dependence well approximated by a linear function of time. 

The \textbf{III region} of pressures starts at the point where an abrupt drop in dislocation velocity is observed. For instance, for sample 2+2 that drop is from the maximum velocity to nearly zero within pressure range of 0.1 GPa. Initially,  for larger samples, at pressures just after that drop, displacement as a  function of time changes abruptly from linear to concave, with slope  decreasing with time, and at certain moment dislocations start to disintegrate. 

Straight lines in left of Fig. \ref{fig:Vscaling} illustrate an attempt to approximate velocity in \textbf{region II} by a linear dependence on pressure, \(V=a \cdot P^0_{xy}+b\). It is found (not shown here) that coefficient \(a\) in this function scales well with the distance \(d\) between the surface at which pressure is applied and dislocations: \(a=26.9 -3.6 \sqrt{d}\) where units of V are in {\AA}/ps while distance \(d\) is the number of unit cells in Y-direction.

We find that a better scaling of velocity in \textbf{region I} and \textbf{region II} is obtained with \(S_{xy}\) instead, as shown in the right of Fig. \ref{fig:Vscaling}. For that an average stress \(S_{xy}\) in the sample space region between the sample bottom and location of dislocations is used. Profiles of \(S_{xy}\) in that region become quasi-stable for all studied samples for times greater than about 30 ps since the time when pressure \(P^0_{xy}\) is applied. For a better accuracy the average has been computed in that entire region for times in interval 40-60 ps.

It ought to be mentioned that although the border line between \textbf{region II} and \textbf{region III} is not fully investigated by us, we think that the pressure when dislocation dissociate depends on what is the position of dislocation core. Equations on displacement field for generating a line dislocation, as for instance these of Carpio and Bonilla \cite{Carpio}, contain a singularity if the core of dislocation is at a position of an atom. Hence, stability of dislocations depends on a proper choice of core position.

\subsection {Velocity scaling with temperature.}

\begin{figure}[ht]
	\centering
	\noindent\includegraphics[scale=1.0]{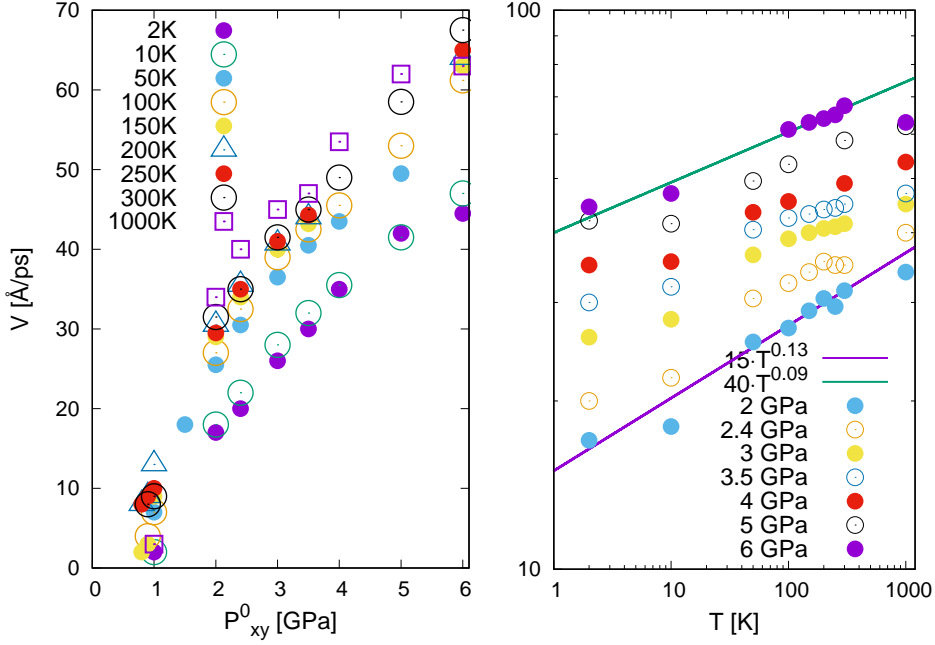}
	\caption{Temperature effect on dislocations velocity. The data on the right in log-log scale are the same as on the left and confirm power-law dependence of velocities on temperature, when measured at constant pressure.}
	\label{velocity_02}
\end{figure}

For a more complete description, in this section we report on temperature scaling of dislocation velocity by using another sample of size 8+8. In this case we obtain results very close to these already presented for T=50K. The difference is that dissociation of dislocations is seen only at applied pressures when dislocation velocity becomes close to the speed of sound (the \textbf{region III} is narrow). On the left of Fig. \ref {velocity_02} velocity is shown as a function of applied pressure for several values of pressure and for a a broad range of temperatures. We see that a critical stress needed for dislocation movement, $\tau_p$ of about 0.75 GPa does not depend on temperature in a noticeable way. This value is in accordance with estimations for edge dislocation in FCC lattice of our geometry, as it is of the order of $10^{-3}-10^{-2}$ of shear modulus, \cite{Kamimura}. An exact value of $\tau_p$ is however difficult for computation and a broad range of theoretical predictions is made \cite{Joos},\cite{Lu}. In the classical Peierls-Nabarro model of stress flow \cite{Hirth}, it is given by:

\begin{equation}
\tau_p=\frac{2G}{1-\nu} exp\left( -\frac{2\pi d}{b(1-\nu)}\right),
\label{eq:peierls}\end{equation}

where $G$ is shear modulus, $\nu$ is Poisson ratio, $b$ is magnitude of Burgers vector and $d$ is distance between glide planes. For Burgers vector $\frac{1}{6}[\bar{2}11]$, $\nu=0.36$, and d=2.06 {\AA} (which is $a/\sqrt{3}$), we obtain $\tau_p/G \approx 0.03$.

Dislocation velocity V(T) is determined by the speed the system can dissipate energy with phonon mediation. This can be seen well in MD simulations, which are of the realm of classical physics: reducing temperature to zero leads to no dislocations movement.
In phenomenological models V(T) is written in the form \cite{Po}, \cite{Yin}:

\begin{equation}
V(T)=\frac{\tau\cdot b}{B(\tau,T)},
\label{eq:tau}\end{equation}

where $\tau$ is stress, $b$ is value of Burgers vector, and $B(\tau,T)$ is approximated by $B_0+B_1T$.

Curves of velocities of dislocations in left of Fig. \ref{velocity_02} as a function of pressure remain similar, change of their slope is small (if any) and it is of about 8{\AA}/($GPa\cdot ps)$), regardless of temperature, albeit these curves shift upward when T increases. This is different from what is expected based on commonly accepted models \cite{Olmsted},\cite{Rodary}, and different than for instance MD results observed for BCC Fe \cite{Queyreau}. It is possible that large stress fluctuations for atoms in crystal lattice, due to atomic inhomogeneity, hinder in our case the effect of thermal fluctuations on dislocations mobility.

The right part of Fig. \ref {velocity_02} shows the same data as on its left side, with the difference that velocities and temperature are in log-log scales. The solid straight lines illustrate a fitting of velocity to a power-law temperature dependence, for a constant value of external pressure applied. 
We find that velocities at constant pressure may be approximated by dependencies of the kind: $V(T) \sim T^{\alpha}$, where ${\alpha}$ is around 0.13 for applied pressure of 2 GPa and it is around 0.09 for 6 GPa. This is in agreement with what to expect based on results reported in the literature. 

Many features of dislocation movement as described here have been reported in the past, but they were not interpreted consistently in terms of dynamic internal shear pressure penetration. For instance, Rodary et al. \cite{Rodary} propose 4-region approximation for velocity vs. pressure diagram, $V-P^0_{xy}$. We do not observe a well defined saturation of velocity for large pressure. Instead dislocations dissociate when velocity is close to sound speed. 
Also, we do not find well defined velocities in \textbf{region I}. Instead we observe in some conditions that movement of dislocations occurs at certain moments only. In \textbf{region II}, which is usually defined as a region of monotonic movement with constant velocity we sometime observe periodic oscillations of velocity with time. These effects are well explained by time-oscillations of pressure profile entering material volume. We find also that instead  of linear $V(P^0_{xy})$ relation in \textbf{region II} a somewhat better approximation of the data would be provided by a power-law dependence of the kind: 
$V=V_0(T)\cdot (P^0_{xy}-\tau_p)^\beta$, where $\tau_p$, a Peierls threshold stress needed for the movement is in our case around 0.75 GPa, and exponent $\beta$ is around 0.7.


\section{Summary.} 
\label{summary}
 
In order to model in molecular dynamics edge (line) dislocations evolution under the local stress field in FCC steel 310S, an analysis of pressure penetration into the sample interior has been carried out. It was shown that pressure (stress field distribution) enters the sample volume with sound speed (of about 70 {\AA}/ps in our case) in a form of waves that can be described well by the damped harmonic oscillator equations, while oscillations frequency of atoms displacements are related to interatomic potential. Angular frequency of these oscillations, of about 29.2/ps, is in agreement with that determined for classical linear harmonic oscillator (for an atom of Fe in a FCC lattice), which we estimate at about 37.0/ps. 

Hence, stress distribution depends on time since the pressure has been applied at the surface and therefore it depends on size of samples. That results on sensitivity of any MD modeling on time scales involved (time step used) and on size of samples studied. Large oscillations of shear surface pressure are are shown to be related to intrinsic material properties and sample size (number of atomic crystallographic planes) along the direction of sound propagation.

Velocity of dislocations scales well with spatially averaged local shear component of virial stress tensor value around dislocation core and not with the external shear pressure.

The maximum dislocation velocity approaches sound speed for high enough stress field. The existence of limiting value of stress (Peierls stress) of about 0.75 GPa is necessary for causing edge dislocations movement. In most cases dislocations move with constant speed after the critical stress is exceeded and that speed is linearly proportional to applied pressure. It is confirmed that dislocation speed dependence on temperature may be described by a power-law formula.


\section*{Acknowledgements}

This work has been supported by the National Science Centre through the Grant No UMO-2020/38/E/ST8/00453.

\section*{References}

\bibliography{virial-els2}

\begin{thebibliography}{10}
\expandafter\ifx\csname url\endcsname\relax
  \def\url#1{\texttt{#1}}\fi
\expandafter\ifx\csname urlprefix\endcsname\relax\def\urlprefix{URL }\fi
\expandafter\ifx\csname href\endcsname\relax
  \def\href#1#2{#2} \def\path#1{#1}\fi

\bibitem{Subramaniyan}
A.~K. Subramaniyan, C.~Sun, Continuum interpretation of virial stress in
  molecular simulations, Int. J. Solids Struct. 45 (2008) 4340--4346.
\newblock \href {http://dx.doi.org/10.1016/j.ijsolstr.2008.03.016}
  {\path{doi:10.1016/j.ijsolstr.2008.03.016}}.

\bibitem{Zimmerman}
J.~Zimmerman, E.~W. III, J.~Hoyt, R.~Jones, P.~Klein, D.~Bammann, Calculation
  of stress in atomistic simulation, Modelling Simul. Mater. Sci. Eng. 12
  (2004) S319.
\newblock \href {http://dx.doi.org/10.1088/0965-0393/12/4/S03}
  {\path{doi:10.1088/0965-0393/12/4/S03}}.

\bibitem{Elder}
R.~M. Elder, W.~D. Mattson, T.~W. Sirk, Origins of error in the localized
  virial stress, Chem. Phys. Lett. 731 (2019) 136580.
\newblock \href {http://dx.doi.org/10.1016/j.cplett.2019.07.008}
  {\path{doi:10.1016/j.cplett.2019.07.008}}.

\bibitem{Yang}
J.~Yang, K.Komvopoulos, A stress analysis method for molecular dynamics
  systems, Int. J. Solids Struct. 193--194 (2020) 98--105.
\newblock \href {http://dx.doi.org/10.1016/j.ijsolstr.2020.02.003}
  {\path{doi:10.1016/j.ijsolstr.2020.02.003}}.

\bibitem{Zhou}
M.~Zhou, A new look at the atomic level virial stress: on continuum-molecular
  system equivalence, Proc. R. Soc. Lond. A 459 (2003) 2347--2392.
\newblock \href {http://dx.doi.org/10.1098/rspa.2003.1127}
  {\path{doi:10.1098/rspa.2003.1127}}.

\bibitem{Simar}
A.~Simar, H.-J.~L. Voigt, B.~D. Wirth, Molecular dynamics simulations of
  dislocation interaction with voids in nickel, Comp. Mat. Sci. 50 (2011)
  1811--1817.
\newblock \href {http://dx.doi.org/10.1016/j.commatsci.2011.01.020}
  {\path{doi:10.1016/j.commatsci.2011.01.020}}.

\bibitem{Kinsler}
L.~Kinsler, A.~Frey, A.~Coppens, J.~Sanders, Fundamentals of Acoustics, 4th
  Edition, John Wiley \& Sons, New York, 2000.

\bibitem{Sprik}
M.~Sprik, R.~W. Impey, M.~L. Klein, Second-order elastic constants for the
  lennard-jones solid, Phys. Rev. B 29 (1984) 4368.
\newblock \href {http://dx.doi.org/10.1103/PhysRevB.29.4368}
  {\path{doi:10.1103/PhysRevB.29.4368}}.

\bibitem{Bonny}
G.~Bonny, N.~Castin, D.~Terentyev, Interatomic potential for studying ageing
  under irradiation in stainless steels: the {FeNiCr} model alloy, Modelling
  Simul. Mater. Sci. Eng. 21~(8) (2013) 085004.
\newblock \href {http://dx.doi.org/10.1088/0965-0393/21/8/085004}
  {\path{doi:10.1088/0965-0393/21/8/085004}}.

\bibitem{Decremps}
F.~Decremps, D.~Antonangeli, M.~Gauthier, S.~Ayrinhac, M.~Morand, G.~L.
  Marchand, F.~Bergame, J.~Philippe, Sound velocity of iron up to 152 {GP}a by
  picosecond acoustics in diamond anvil cell, Geophys. Res. Lett. 41 (2014)
  1459--1464.
\newblock \href {http://dx.doi.org/10.1002/2013GL058859}
  {\path{doi:10.1002/2013GL058859}}.

\bibitem{LAMMPS}
S.~Plimpton, Fast parallel algorithms for short-range molecular dynamics, J.
  Comp. Phys. 117 (1995) 1.
\newblock \href {http://dx.doi.org/10.1006/jcph.1995.1039}
  {\path{doi:10.1006/jcph.1995.1039}}.

\bibitem{Hayek}
S.~I. Hayek, Mechanical vibration and damping, in: Encyclopedia of Applied
  Physics, WILEY-VCH Verlag GmbH \& Co KGaA, 2003.
\newblock \href {http://dx.doi.org/10.1002/3527600434.eap231}
  {\path{doi:10.1002/3527600434.eap231}}.

\bibitem{Thornton}
S.~Thornton, J.~Marion, Classical Dynamics of Particles and Systems, 5th
  Edition, Brooks Cole, Pacific Grove, CA, 2003.

\bibitem{Monnet}
G.~Monnet, D.~Terentyev, Structure and mobility of the 1/2$<$111$>$\{112\} edge
  dislocation in bcc iron studied by molecular dynamics, Acta Materialia 57
  (2009) 1416--1426.
\newblock \href {http://dx.doi.org/10.1016/j.actamat.2008.11.030}
  {\path{doi:10.1016/j.actamat.2008.11.030}}.

\bibitem{Atomsk}
P.~Hirel, Atomsk: A tool for manipulating and converting atomic data files,
  Comput. Phys. Comm. 197 (2015) 212.
\newblock \href {http://dx.doi.org/10.1016/j.cpc.2015.07.012}
  {\path{doi:10.1016/j.cpc.2015.07.012}}.

\bibitem{Artur}
L.~Béland, A.~Tamm, S.~Mu, G.~Samolyuk, Y.~Osetsky, A.~Aabloo, M.~Klintenberg,
  A.~Caro, R.~Stoller, Accurate classical short-range forces for the study of
  collision cascades in {Fe-Ni-Cr}, Comput. Phys. Comm. 219 (2017) 11--19.
\newblock \href {http://dx.doi.org/10.1016/j.cpc.2017.05.001}
  {\path{doi:10.1016/j.cpc.2017.05.001}}.

\bibitem{Mendelev}
M.~Mendelev, to be published (2019).
\newblock \href {http://dx.doi.org/10.25950/538764d4}
  {\path{doi:10.25950/538764d4}}.

\bibitem{ZhouXW}
X.~Zhou, M.~Foster, R.~Sills, An {Fe-Ni-Cr} embedded atom method potential for
  austenitic and ferritic systems, J. Comput. Chem. 39~(29) (2018) 2420--2431.
\newblock \href {http://dx.doi.org/10.1002/jcc.25573}
  {\path{doi:10.1002/jcc.25573}}.

\bibitem{Wu}
C.~Wu, B.-J. Lee, X.~Su, Modified embedded-atom interatomic potential for
  {Fe-Ni}, {Cr-Ni} and {Fe-Cr-Ni} systems, Calphad 57 (2017) 98--106.
\newblock \href {http://dx.doi.org/10.1016/j.calphad.2017.03.007}
  {\path{doi:10.1016/j.calphad.2017.03.007}}.

\bibitem{Osetsky}
S.~Zhao, Y.~N. Osetsky, Y.~Zhang, Atomic-scale dynamics of edge dislocations in
  {Ni} and concentrated solid solution {NiFe} alloys, J. Alloy Comp. 701 (2017)
  1003--1008.
\newblock \href {http://dx.doi.org/10.1016/j.jallcom.2017.01.165}
  {\path{doi:10.1016/j.jallcom.2017.01.165}}.

\bibitem{Carpio}
A.~Carpio, L.~L. Bonilla, Atomic models of dislocations and their motion in
  cubic crystals, in: P.~Neittaanmaki, T.~Rossi, S.~Korotov, E.~Onate,
  J.~Periaux, D.~Knorzer (Eds.), European Congress on Computational Methods in
  Applied Sciences and Engineering ECCOMAS 2004, 2004.

\bibitem{zhang}
J.-Y. Zhang, W.-Z. Zhang, A general method to construct dislocations in
  atomistic simulations, Modelling Simul. Mater. Sci. Eng. 27~(3) (2019)
  035008.
\newblock \href {http://dx.doi.org/10.1088/1361-651X/ab021a}
  {\path{doi:10.1088/1361-651X/ab021a}}.

\bibitem{Nabarro}
W.~Cai, V.~V. Bulatov, J.~Chang, J.~Li, S.~Yip, Dislocation core effects on
  mobility, in: F.~Nabarro, J.~Hirth (Eds.), Dislocations in Solids, Vol.~12,
  North-Holland, 2004, p.~1.

\bibitem{Ovito}
A.~Stukowski, Visualization and analysis of atomistic simulation data with
  {OVITO} - the {O}pen {V}isualization {T}ool, Modelling Simul. Mater. Sci.
  Eng. 18~(1) (2010) 015012.
\newblock \href {http://dx.doi.org/10.1088/0965-0393/18/1/015012}
  {\path{doi:10.1088/0965-0393/18/1/015012}}.

\bibitem{CAT}
A.~Stukowski, Structure identification methods for atomistic simulations of
  crystalline materials, Modelling Simul. Mater. Sci. Eng. 20 (2012) 045021.
\newblock \href {http://dx.doi.org/10.1088/0965-0393/20/4/045021}
  {\path{doi:10.1088/0965-0393/20/4/045021}}.

\bibitem{Lui}
L.~C. H, H.~T. F, Measurement of layer breathing mode vibrations in few-layer
  graphene, Phys. Rev. B 87 (2013) 121404.
\newblock \href {http://dx.doi.org/10.1103/PhysRevB.87.121404}
  {\path{doi:10.1103/PhysRevB.87.121404}}.

\bibitem{Tan}
P.~Tan, W.~Han, W.J, Zhao, Z.~Wu, K.~Chang, H.~Wang, Y.~Wang, N.~Bonini,
  N.~Marzari, G.~Savini, A.~Lombardo, A.~Ferrari, The shear mode of multilayer
  graphene, Nat. Mat. 11 (2012) 294.
\newblock \href {http://dx.doi.org/10.1038/nmat3245}
  {\path{doi:10.1038/nmat3245}}.

\bibitem{Koziol}
Z.~Kozio{\l}, G.~Gawlik, J.~Jagielski, Van der waals interlayer potential of
  graphitic structures: From {L}ennard-{J}ones to {K}olmogorov-{C}respy and
  {L}ebedeva models, Chin. Phys. B 28 (2019) 096101.
\newblock \href {http://dx.doi.org/10.1088/1674-1056/ab38a5}
  {\path{doi:10.1088/1674-1056/ab38a5}}.

\bibitem{Muller}
M.~Müller, P.~Erhart, K.~Albe, Analytic bond-order potential for bcc and fcc
  iron-comparison with established embedded-atom method potentials, J. Phys.:
  Condens. Matter 19 (2007) 326220.
\newblock \href {http://dx.doi.org/10.1088/0953-8984/19/32/326220}
  {\path{doi:10.1088/0953-8984/19/32/326220}}.

\bibitem{Zarestky}
J.~Zarestky, C.~Stassis, Lattice dynamics of $\gamma$-{Fe}, Phys. Rev. B 35
  (1987) 4500.
\newblock \href {http://dx.doi.org/10.1103/PhysRevB.35.4500}
  {\path{doi:10.1103/PhysRevB.35.4500}}.

\bibitem{Stankov}
S.~Stankov, R.~Röhlsberger, T.~Ślęzak, M.~Sladecek, B.~Sepiol, G.~Vogl,
  A.~I. Chumakov, R.~Rüffer, N.~Spiridis, J.~Łażewski, K.~Parliński,
  J.~Korecki, Phonons in iron: From the bulk to an epitaxial monolayer, Phys.
  Rev. Lett. 99 (2007) 185501.
\newblock \href {http://dx.doi.org/10.1103/PhysRevLett.99.185501}
  {\path{doi:10.1103/PhysRevLett.99.185501}}.

\bibitem{Danilkin}
S.~A. Danilkin, H.~Fuess, T.~Wieder, Phonon dispersion and elastic constants in
  {Fe-Cr-Mn-Ni} austenitic steel, J. Mater. Sci. 36 (2001) 811.
\newblock \href {http://dx.doi.org/10.1023/A:1004801823614}
  {\path{doi:10.1023/A:1004801823614}}.

\bibitem{Urbassek}
H.~Urbassek, L.~Sandoval, Diffusionless transformations high strength steels
  modelling and advanced analytical techniques, in: E.~Pereloma, D.~V. Edmonds
  (Eds.), Phase Transformations in Steels, Woodhead Publishing, 2012.

\bibitem{Thomson}
A.~P. Thompson, S.~J. Plimpton, W.~Mattson, General formulation of pressure and
  stress tensor for arbitrary many-body interaction potentials under periodic
  boundary conditions, J. Chem. Phys. 131 (2009) 154107.
\newblock \href {http://dx.doi.org/10.1063/1.3245303}
  {\path{doi:10.1063/1.3245303}}.

\bibitem{Kamimura}
Y.~Kamimura, K.~Edagawa, S.~Takeuchi, Experimental evaluation of the peierls
  stresses in a variety of crystals and their relation to the crystal
  structure, Acta Materialia 61 (2013) 294--309.
\newblock \href {http://dx.doi.org/10.1016/j.actamat.2012.09.059}
  {\path{doi:10.1016/j.actamat.2012.09.059}}.

\bibitem{Joos}
B.~Jo{\'o}s, M.~S. Duesbery, The peierls stress of dislocations: An analytic
  formula, Phys. Rev. Lett. 78 (1997) 266--269.
\newblock \href {http://dx.doi.org/10.1103/PhysRevLett.78.266}
  {\path{doi:10.1103/PhysRevLett.78.266}}.

\bibitem{Lu}
G.~Lu, The peierls-nabarro model of dislocations: A venerable theory and its
  current development, in: S.~Yip (Ed.), Handbook of Materials Modeling. Volume
  I: Methods and Models, Springer, 2005, pp. 1--19.
\newblock \href {http://dx.doi.org/10.1007/978-1-4020-3286-8_41}
  {\path{doi:10.1007/978-1-4020-3286-8_41}}.

\bibitem{Hirth}
J.~Hirth, J.~Lothe, Theory of Dislocations, 2nd Edition, John Wiley \& Sons,
  New York, 1992.

\bibitem{Po}
G.~Po, Y.~Cui, D.~Rivera, D.~Cereceda, T.~D. Swinburne, J.~Marian, N.~Ghoniem,
  A phenomenological dislocation mobility law for bcc metals, Acta Materialia
  119 (2016) 123--135.
\newblock \href {http://dx.doi.org/10.1016/j.actamat.2016.08.016}
  {\path{doi:10.1016/j.actamat.2016.08.016}}.

\bibitem{Yin}
S.~Yin, Y.~Zuo, A.~Abu-Odeh, H.~Zheng, S.~P. Ong, M.~Asta, R.~O. Ritchie,
  Atomistic simulations of dislocation mobility in refractory high-entropy
  alloys and the effect of chemical short-range order, Nat. Comm. 12 (2021)
  4873.
\newblock \href {http://dx.doi.org/10.1038/s41467-021-25134-0}
  {\path{doi:10.1038/s41467-021-25134-0}}.

\bibitem{Olmsted}
D.~L. Olmsted, J.~Louis G.~Hector, W.~A. Curtin, R.~J. Clifton, Atomistic
  simulations of dislocation mobility in {Al}, {Ni} and {Al/Mg} alloys,
  Modelling Simul. Mater. Sci. Eng. 13 (2005) 371.
\newblock \href {http://dx.doi.org/10.1088/0965-0393/13/3/007}
  {\path{doi:10.1088/0965-0393/13/3/007}}.

\bibitem{Rodary}
E.~Rodary, D.~Rodney, L.~Proville, Y.~Br\'echet, G.~Martin, Dislocation glide
  in model {N}i({A}l) solid solutions by molecular dynamics, Phys. Rev. B 70
  (2004) 054111.
\newblock \href {http://dx.doi.org/10.1103/PhysRevB.70.054111}
  {\path{doi:10.1103/PhysRevB.70.054111}}.

\bibitem{Queyreau}
S.~Queyreau, J.~Marian, M.~R. Gilbert, B.~D. Wirth, Edge dislocation mobilities
  in bcc fe obtained by molecular dynamics, Phys. Rev. B 84 (2011) 064106.
\newblock \href {http://dx.doi.org/10.1103/PhysRevB.84.064106}
  {\path{doi:10.1103/PhysRevB.84.064106}}.

\end{thebibliography}

\end{document}